\documentclass[twocolumn,,amsmath,amssymb,prb,aps,superbib,tightenlines,floatfix]{revtex4-1}

\usepackage{graphicx,graphicx,epsfig}
\usepackage{hyperref}

\begin{document}

\title{Memory circuit elements: from systems to applications}

\author{Yu. V. Pershin$^{1,}$\footnote{Corresponding author, email:pershin@physics.sc.edu.}, J. Martinez-Rincon$^1$, M. Di Ventra$^2$}
\affiliation{$^1$Department of Physics and Astronomy and USC
Nanocenter, University of South Carolina, Columbia, SC, 29208 USA
\\ $^2$Department
of Physics, University of California, San Diego, La Jolla, CA
92093-0319, USA}

\begin{abstract}
In this paper, we briefly review the concept of memory circuit elements, namely memristors, memcapacitors and meminductors, and then discuss some applications by focusing
mainly on the first class. We present several examples,
their modeling and applications ranging from analog programming to biological systems. Since the phenomena associated with memory are ubiquitous at the nanoscale,
we expect the interest in these circuit elements to increase in coming years.
\end{abstract}


\maketitle

\section{Introduction}

Currently, there is strong interest in circuit elements with
history-dependent properties. These form a general class of memory elements~\cite{diventra09a} which includes memory resistors (memristors)~\cite{chua71a}, memory
capacitors (memcapacitors) and memory inductors (meminductors).
Quite generally, memristors, memcapacitors and meminductors can be
considered as time- and history-dependent generalizations of their standard
counterparts.

Mathematically, an $n$th-order $u$-controlled memory element is
defined by the equations~\cite{diventra09a}
\begin{eqnarray}
y(t)&=&g\left(x,u,t \right)u(t) \label{Geq1}\\ \dot{x}&=&f\left(
x,u,t\right) \label{Geq2}.
\end{eqnarray}
Here, $u(t)$ and $y(t)$ are any two circuit variables (current,
charge, voltage, or flux) denoting input and output of the system,
$x$ is an $n$-dimensional vector of internal state variables, $g$ is
a generalized response, and $f$ is a continuous $n$-dimensional
vector function. Special interest is devoted to devices determined
by three pairs of circuit variables: current-voltage (memristors),
charge-voltage (memcapacitors), and flux-current (meminductors).
Two other pairs (charge-current and voltage-flux) are linked
through equations of electrodynamics, and therefore do not give rise to any new
device. Devices defined by the relation of charge and
flux (which is the integral of the voltage) are not considered as
a separate group since such devices can be redefined in the
current-voltage basis~\cite{chua71a}.

We stress first that these memory devices are
{\em dynamical} systems, namely systems whose state and response
may change in time. In typical applications, memory devices operate under the
action of a time-dependent (not static) input. It is also worth
noting that the relations~(\ref{Geq1}) and~(\ref{Geq2}) are quite general.
In particular, they may define
systems whose properties {\em cannot} be reproduced with standard
resistors, capacitors and inductors. In other words, there is no
possible combination of standard circuit elements that can
reproduce the dynamical properties of certain memory elements. In
this sense, memristors, memcapacitors and meminductors can be
considered as ``fundamental'' circuit elements. However, at least
in the case of memcapacitors and meminductors, the
relations~(\ref{Geq1}) and~(\ref{Geq2}) may also represent some other
systems whose dynamical properties can be simulated by an
appropriate combination of standard (albeit possibly non-linear)
elements (for examples of these, see, e.g.,
Refs.~\onlinecite{krems2010a,martinez09a}). Nonetheless, these
latter memory systems are still of great importance since they
provide a complex functionality within a single electronic
structure~\cite{martinez09a}.

The area of research of memristors is
more advanced both technologically and theoretically than that of
memcapacitors and meminductors. In fact, although some experimental
systems have been identified as memcapacitive~\cite{diventra09a,krems2010a,martinez09a,driscoll09a,Lai09a} and
meminductive~\cite{diventra09a}, their number is
still small and possible applications for them are less developed.
In this review, we will then focus on memristors
\footnote{According to existing terminology, we will use the
term memristors for both ideal memristors~\cite{chua71a} and
memristive systems~\cite{chua76a}.}.

It is convenient to introduce voltage-controlled and
current-controlled memristors~\cite{chua76a}. An $n$th-order
current-controlled memristive system is described by the equations
\begin{eqnarray}
V_M(t)&=&R\left(x,I,t \right)I(t) \label{eq1}\\
\dot{x}&=&f\left(x,I,t\right) \label{eq2}
\end{eqnarray}
with $x$ again a vector representing $n$ internal state variables,
$V_M(t)$ and $I(t)$ denote the voltage and current across the
device, and $R$ is a scalar, called the {\em memristance} (for
memory resistance). The equation for a charge-controlled (or {\it
ideal}) memristor is a particular case of Eqs. (\ref{eq1}) and
(\ref{eq2}), when $R$ depends only on charge, namely

\begin{equation}
V_M=R\left( q\left(t \right)\right)I, \label{eq3}
\end{equation}
with the charge related to the current via time derivative:
$I=dq/dt$.

We can also define an $n$th-order {\it voltage}-controlled
memristive system from the relations
\begin{eqnarray}
I(t)&=&G\left(x,V_M,t \right)V_M(t) \label{Condeq1}\\
\dot{x}&=&f\left( x,V_M,t\right) \label{Condeq2}
\end{eqnarray}
where we call $G$ the {\it memductance} (for memory conductance).

\begin{figure}[tb]
\centering
\includegraphics[width=7cm]{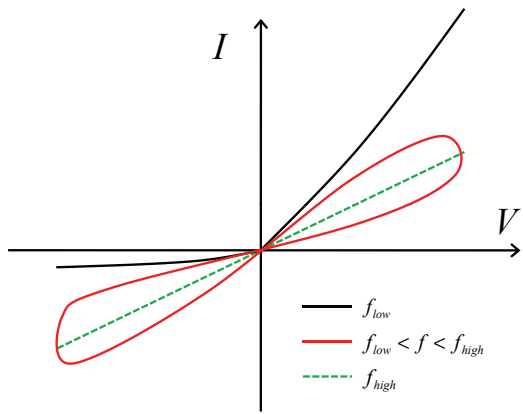}
\caption{Schematic of typical $I-V$ curves of a memristor obtained when an ac-voltage of frequency $f$ is applied.
At low frequencies $f_{low}$, the $I-V$ curve is a non-linear curve, at intermediate frequencies $f_{low}<f<f_{high}$, it is a pinched hysteresis loop, and at high frequencies $f_{high}$, it is a straight line. Note that all
curves pass through (0,0).
\label{fig:hysteresis}}
\end{figure}

There are several common properties of
memristors~\cite{chua71a,chua76a,diventra09a}. They are passive
devices (provided $R\geq 0$ or $G\geq 0$) without an energy
storage capability (unlike memcapacitors and meminductors which
can store energy). Typically, at high frequencies, they behave as
linear resistors and at low frequencies as non-linear resistors
(assuming existence of a steady-state solution of Eq. (\ref{eq2})
or (\ref{Condeq2})). A distinctive feature of well-defined
memristors (with non-zero $R$ or $G$) is a pinched hysteresis loop
on the current-voltage plane (see Fig. \ref{fig:hysteresis}).

Because of the short format of this review, we are not able to
include all topics concerning memristors or to cover the ones
included in due detail. We will focus mainly on our results in
this field and organize the text in the following order. Sec.
\ref{Sec2} describes several different realizations of memristors.
A theoretical modeling
is discussed in Sec. \ref{Sec3}. We overview several important
applications of memristors in Sec. \ref{Sec4}. Finally, Sec.
\ref{Sec5} presents our conclusions.

\section{Experimental Realizations \label{Sec2}}

Many different systems exhibit memristive behavior.
Early-identified memristor's examples~\cite{chua76a} include
thermistors~\cite{sapoff63a} and ionic systems~\cite{hodgkin52a}.
Recent interest in memristor's realizations has centered on
nanoscale systems and nanostructures as we discuss below. Some of
them were intensively studied in the context of development of
resistive-switching memory during last 10-15 years. This area, to
the best of our knowledge, was initiated by Hickmott in 1962 by the
observation of hysteretic behavior in oxide
insulators~\cite{hickmott62a}.

At the present time, it is well known that memristive effects can
be observed in such diverse classes of materials as binary oxides
(TiO$_2$, CuO, NiO, CoO, Fe$_2$O$_3$, MoO,
VO$_2$)~\cite{yang08a,inoue08a,lee07a,seo04a,driscoll09a,driscoll09b},
perovskite-type oxides (Pr$_{1-x}$Ca$_x$MnO$_3$,
SrTiO$_3$:Cr)~\cite{asamitsu97a,fors05a,kim06a,meijer05a,nian07a},
sulfides (Cu$_2$S,Ag$_2$S)~\cite{terabe05a,tamura06a,waser07a},
semiconductors (Si, GaAs, ZnSe-Ge)~\cite{jo08a,dong08a,jo09a},
spintronics materials~\cite{pershin08a,pershin09a,wang09a} and
organics~\cite{stewart04a,lai05a,alibart10a}. In this review, we
limit our discussion to the binary oxide
TiO$_2$ and spintronics systems. We will also discuss a memristor
emulator~\cite{pershin09d,pershin09c} - an electronic module simulating the
memristor behavior.

\subsection{Binary oxides materials}

TiO$_2$ is an example of binary oxide showing bipolar resistive
switching. Recently, a memristive mechanism of its switching
behavior has been identified~\cite{strukov08a,yang08a}. It was
suggested that the switching involves changes to the electronic
barrier at the Pt/TiO$_2$ interface induced by drift of oxygen vacancies
under an applied electric field. When vacancies drift toward
the interface, they create conducting channels that short-circuit
the electronic barrier. When vacancies drift away from the interface they
eliminate these channels, restoring the original electronic
barrier~\cite{yang08a}. This material demonstrates promising
characteristics for future ultra-high density memory applications
such as fast read/write times ($\sim$10ns), high on/off ratios
($\sim 10^3$), suitable range of programming voltages, and
possibility to fabricate small size cells.

\subsection{Spintronic materials}

\begin{figure}[tb]
\centering
\includegraphics[width=7cm]{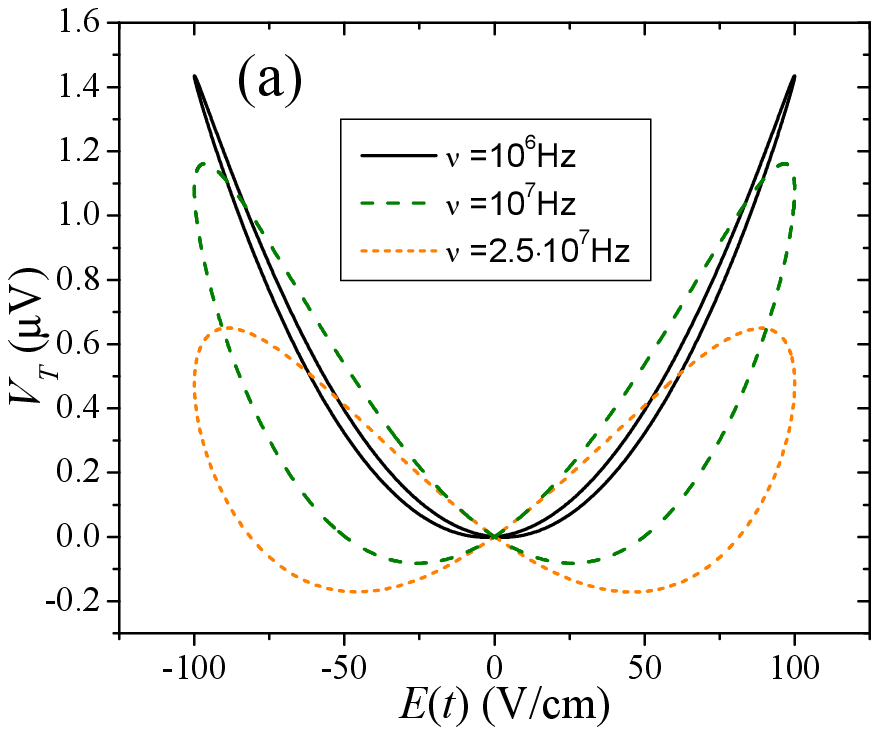}
\includegraphics[width=7cm]{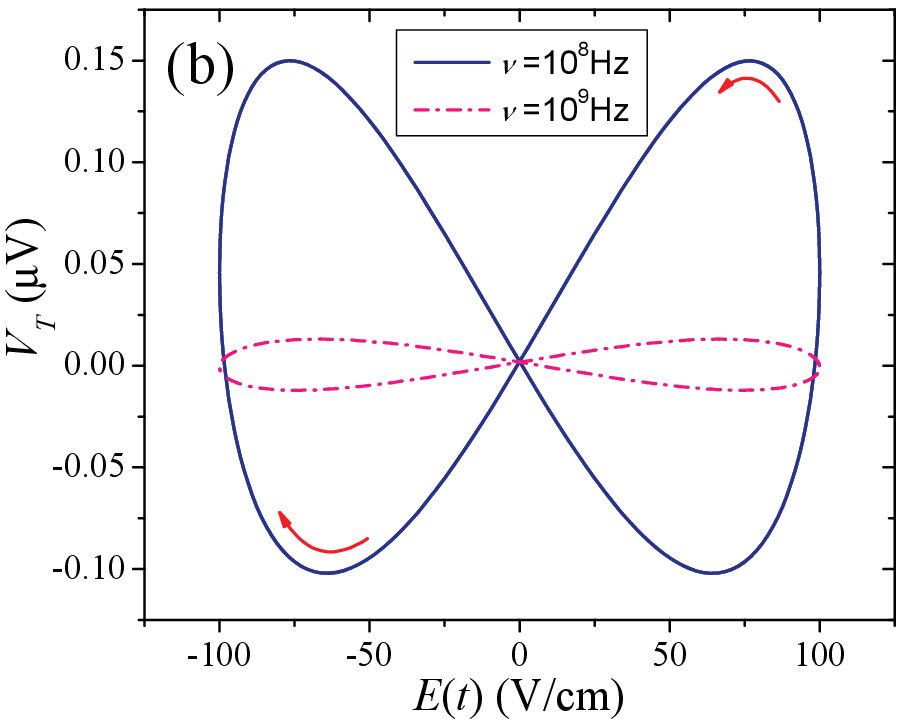}
\caption{Spin memristive effects in a semiconducting system with
inhomogeneous electron density in the direction perpendicular to
main current flow~\cite{pershin09a}. Here, we show the transverse
voltage as a function of applied electric field at different
applied field frequencies as indicated. (From
Ref.~\onlinecite{pershin09a} with permission. Copyright American
Physical Society, 2009).} \label{fig:spintransverse}
\end{figure}

Semiconductor spintronic systems~\cite{zutic04a} were identified
as memristive systems by Pershin and Di Ventra~\cite{pershin08a}.
In semiconductor spintronics, memory effects are completely due to
the electron spin degree of freedom. When an external control
parameter changes, it takes time for the electron spin polarization to
adjust to a new control parameter value (typically, the
equilibration is governed by electron-spin diffusion and
relaxation processes). Therefore, if a system's resistivity
depends on the level of electron spin polarization (as in
semiconductor-ferromagnet junctions~\cite{pershin07a,pershin08b})
then such a system is fundamentally memristive~\cite{pershin08a}. Moreover,
memory effects were predicted for spin Hall effect systems with inhomogeneous charge density distribution transverse to the driving
electric field~\cite{pershin09a}. It is interesting that in this case memory
effects can be observed directly in the transverse voltage. Fig.
\ref{fig:spintransverse} shows transverse voltage hysteresis loops
that demonstrate typical memristive behavior: non-linear
dependence at low frequencies, pronounced hysteresis at higher
frequencies and hysteresis collapse at very high frequencies~\cite{pershin09a}.

In metal spintronics, spin torque transfer known for a long
time~\cite{berger78a,slonczewski96a,tsoi98a,myers99a} can also be categorized as
memristive. In spin torque transfer systems, the resistance
is determined by the relative magnetization between opposite sides
of a magnetic tunnel junction. Current flowing through the
junction induces spin torque, in turn changing the relative
magnetization. Krzysteczko et al.~\cite{krzysteczko09a} fabricated
a modified magnetic tunnel junction with a resistive switching
material inside. A spintronic memristor based on
spin-torque-induced magnetization motion was discussed by Wang {\it et
al.}~\cite{wang09a}.

\subsection{Memristor emulator}

A memristor emulator is an electronic module composed of standard (off-the-shelf)
electronic components whose circuit response imitates that of
memristor. Memristor emulator consisting of a
microcontroller-controlled digital potentiometer and an
analog-to-digital converter was recently demonstrated~\cite{pershin09d,pershin09c}. Using the analog-to-digital converter,
the microcontroller obtains information about the voltage applied to
the memristor emulator and updates the digital potentiometer
resistance using equations of voltage-controlled (Eqs.
(\ref{Condeq1},\ref{Condeq2})) or current-controlled (Eqs.
(\ref{eq1},\ref{eq2})) memristive systems. From the point of view
of the external circuit, the memristor emulator is then a black box
with two connections that operates exactly as a memristor. Its low
cost, simplicity and flexibility make it useful for building
prototype circuits with memrsistors for both research and
education. Moreover, using the memristor
emulator (which could be replaced by a real memristor), modules showing memcapacitive or meminductive behavior
have also been demonstrated~\cite{pershin09e}.

\section{Theoretical description \label{Sec3}}

For practical use of memory elements, the ability to predict their
dynamics in electronic circuits is important. For this purpose, it
is necessary to know the functions $g(x,u,t)$ and $f(x,u,t)$
entering Eqs. (\ref{Geq1},\ref{Geq2}) that can be later used, for
example, in SPICE
modeling~\cite{Benderli2009-1,Biolek2009-1,Biolek2009-2}.
Generally, these functions can be constructed phenomenologically
or derived from an appropriate physical analysis. Although models
of resistive-switching devices were suggested before
2008~\cite{rozenberg04a}, the realization that these devices
belong to the class of memristive systems \cite{strukov08a} has
spurred further progress in understanding their
behavior~\cite{strukov09a, strukov09b,joglekar09a}. Below, we
consider a phenomenological model of memristor suggested by two of
us (YVP and MD) to explain the learning properties of unicellular
organisms~\cite{pershin09b} and a computationally simple tunneling
barrier model. The tunneling barrier model includes both
activation-type of internal state variables dynamics and electron
tunneling through the barrier and it is indeed a physical
realization of the above phenomenological model. It is also useful
to understand the memory resistive properties of several systems
of present interest.

\subsection{Phenomenological model}

A phenomenological model of a memristor has been suggested in Ref.
\onlinecite{pershin09b}. This model captures two very important
experimental features of TiO$_2$ memristors\cite{yang08a}: {\it
i}) the fact that the resistance changes between two limiting
values $R_{min}$ and $R_{max}$, and {\it ii}) a threshold-type
resistance dynamics. The latter is manifested in the following
experimental observation~\cite{yang08a}: at low applied voltages
the resistance of the memristor changes slowly while at high
applied voltages its change is fast. The phenomenological
model describes a voltage-controlled memristor
(see Eqs. (\ref{Condeq1}, \ref{Condeq2})) with
 $R_M=x$ in Eq. (\ref{Condeq1}) and Eq. (\ref{Condeq2}) written as (with
 the memristance acquiring the limiting values $R_{min}$ and $R_{max}$)~\cite{pershin09b}

\begin{figure*}[tb]
\centering
\includegraphics[width=12cm]{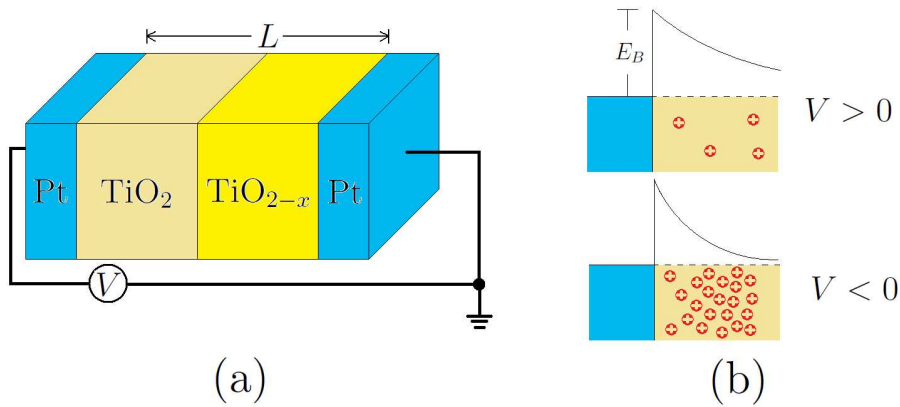}
\caption{Schematic of TiO$_2$ memristor (a) and of its resistivity
switching mechanism (b). The Schottky barrier is reduced in
reverse bias situation when oxygen vacancies are shifted to the
left. \label{fig:cartoon}}
\end{figure*}

\begin{eqnarray}
\dot x&=&\left(\beta V_M+0.5\left( \alpha-\beta\right)\left[
|V_M+V_T|-|V_M-V_T| \right]\right) \nonumber\\ & &\times
\theta\left( x-R_{min}\right) \theta\left( R_{max}-x\right) \;
\label{Mmodel2},
\end{eqnarray}
where $\alpha$ and $\beta$ are constants defining the memristance rate
of change below and above the threshold voltage $V_T$; $V_M$ is
the voltage across the memristor, and $\theta(\cdot)$ is the step function. This model has been used in
several cases ranging from modeling of learning to emulators~\cite{pershin09b,pershin09d,pershin09e}. Moreover, similar models can be applied to memcapacitors and meminductors~\cite{diventra09a}.

\subsection{Tunneling Barrier Model}

Quite generally, let us assume that the electron transport through a voltage-controlled memristor involves a quantum mechanical tunneling through a barrier whose width and height are determined by internal state variables of the device.
A particular example of application of this model regards the Pt/TiO$_2$/Pt thin-film memristor~\cite{yang08a}, in which a Schottky barrier plays the role of the tunneling barrier and the internal state variable is the density of oxygen vacancies in the Schottky barrier region. For the sake of clarity we then focus on this particular system. We also note that the Schottky barrier was not explicitly included in previous titanium dioxide memristor models~\cite{strukov08a,strukov09b}.

\subsubsection{Dynamics of Internal State Variables}

The structure of a TiO$_2$ memristor is very
simple~\cite{yang08a}, as shown in Fig. \ref{fig:cartoon}(a). It
consists of two metallic (e.g., platinum) electrodes separated by
a titanium dioxide thin film containing a region with a high
concentration of oxygen vacancies, that we denote with
TiO$_{2-x}$, with $x$ describing the vacancies concentration. The
right contact in Fig. \ref{fig:cartoon}(a) of the TiO$_{2-x}$
region with Pt can be considered as ohmic~\cite{yang08a}. The left
contact (TiO$_{2}$ with Pt) involves a Schottky
barrier~\cite{yang08a}.

A fraction of oxygen vacancies can be moved from the TiO$_{2-x}$
region to the rest of TiO$_2$ by an applied electric field via the
phenomenon of electro-migration (see, e.g., Ref.
\onlinecite{Maxbook}). Let us consider only mobile vacancies and
track only the total number of mobile vacancies in the Schottky
barrier region that extends inside of TiO$_2$.
Moreover, we assume that in our device the separation between Pt
electrodes is of the order of two Schottky barrier lengths (here,
by Schottky barrier length we mean its length when the
device is in its high-resistance state). Only the vacancies in the
Schottky barrier region are important to us since they modify the
shape of the barrier (this is schematically shown in Fig.
\ref{fig:cartoon}(b)) influencing directly the electron transport
through the device~\footnote{A more detailed modeling of vacancy
dynamics is possible using diffusion equations~\cite{strukov09b}.
However, in this case, simulations become much more
computationally expensive.}.

The vacancy dynamics can be described by a rate equation
\begin{equation}
\frac{dN_L}{dt}=\frac{1}{\tau}\left[-\theta(V)N_L+\theta(-V)N_R\right],
\label{eq:N}
\end{equation}
where $V$ is the applied voltage and $N_{L,R}$ are the number of
mobile vacancies in the left $L$ (Schottky barrier) region and
right, $R$, remaining part of titanium dioxide. The total number $N$
of mobile vacancies is assumed to be constant ($N=N_L+N_R$). At
the initial moment of time, $N_L$ can take any value between $0$
and $N$ depending on the initial state of the system. Limiting cases $N_L=0$ and
$N_L=N$ correspond to high-resistance and low-resistance states of
the memristor, respectively.

The time $\tau$ taken for a vacancy to move from one region to
another can be found using an activation-type expression for the
drift velocity as proposed in Ref.
\onlinecite{strukov09a}
\begin{eqnarray}
\tau&\approx&\frac{L/2}{|v|}\approx\frac{L}{2\,\mu\,E_0\,\sinh\left(|E|/E_0\right)}\approx\frac{L}{2\,\mu\,E_0\,\sinh\left(|V|/L\,E_0\right)}\nonumber\\
&\approx&\left\{\begin{matrix}\frac{L^2}{2\,\mu\,|V|}\hspace{1cm}\textrm{if}\;|V|\ll L\,E_0\\\frac{L}{\mu\,E_0}\exp\left(\frac{-|V|}{L\,E_0}\right)\hspace{1cm}\textrm{if}\;|V|\gg LE_0\end{matrix}\right.,  \label{eq:tau}
\end{eqnarray}
where $\mu$ is the ion mobility, $E_0=2k_BT/(qa)$, $T$ is the
temperature, $q$ is the elementary charge, $a$ is the potential period,
and for the electric field $E$ we have used the
approximation $|E|=|V|/L$, where $L$ is the length of titanium
dioxide material. The derivation of Eq. (\ref{eq:tau}) is based on
the assumption that the oxygen vacancies are confined by a
periodic potential due to all other atoms in the solid along the migration path. The
application of the external electric field deforms the potential
reducing the potential barrier between adjacent potential
wells~\cite{strukov09a}. As a result, the expression for the drift
velocity becomes very non-linear at strong applied
fields~\cite{strukov09a}. Using Eq. (\ref{eq:tau}), Eq.
(\ref{eq:N}) can be rewritten as
\begin{equation}
\label{vacandynamics} \frac{dn_L}{dt}=\frac{2\mu
E_0}{L}\,\sinh\left[|V|/(LE_0)\right]\,\left[-\theta(V)n_L+\theta(-V)\left\{1-n_L\right\}\right],
\end{equation}
where $n_L=N_L/N$. A numerical solution of Eq.
(\ref{vacandynamics}) is shown in Fig.
\ref{fig:devicesimulation}a. Later, we use $n_L(t)$ as an input
parameter for tunneling current calculations.

\subsubsection{Electron transport}

Experimental data~\cite{yang08a} suggest that the tunneling
current through the Schottky barrier is the main limiting factor
to electron transport. This current can be approximated by a
standard expression~\cite{padovani66a}
\begin{equation}
j=\frac{A e^{-b_1}}{\left( c_1 k T\right)^2} \frac{\pi c_1 k
T}{\sin\left( \pi c_1 kT \right)}\left[ 1-e^{-c_1qV}\right],
\label{eq:J}
\end{equation}
where $A=4\pi m^* q (k_BT)^2/h^3$ is the classical Richardson
constant, $m^*$ is the effective electron mass, $b_1$ and $c_1$
are constants. The expressions for $b_1$ and $c_1$ are different
for forward and reverse biases. In particular, in the forward bias
when electrons tunnel from TiO$_2$ into the metal (expressions for
$b_1$ and $c_1$ for the case of reverse bias can be found in Ref.
\onlinecite{padovani66a}),
\begin{eqnarray}
b_1=\frac{E_B-qV}{E_{00}} \\ c_1=\frac{1}{2E_{00}}\ln\left[
\frac{4(E_B-qV)}{\xi}\right]
\end{eqnarray}
where $E_B$ is the barrier height,
difference between work function of the metal and the affinity of the
semiconductor, $\xi$ is the Fermi energy of the semiconductor and
the energy $E_{00}$ is a material-dependent parameter that is
inversely proportional to the Schottky barrier
width~\cite{padovani66a}. For the sake of simplicity, a constant
barrier height is assumed in our calculations. The tunneling through the barrier
dominates thermionic emission and thermionic-field emission
mechanisms when $E_{00}\gg kT$ (the regime we consider here)~\cite{sze_book1}.

\begin{figure}[tb]
\centering
\includegraphics[width=7cm]{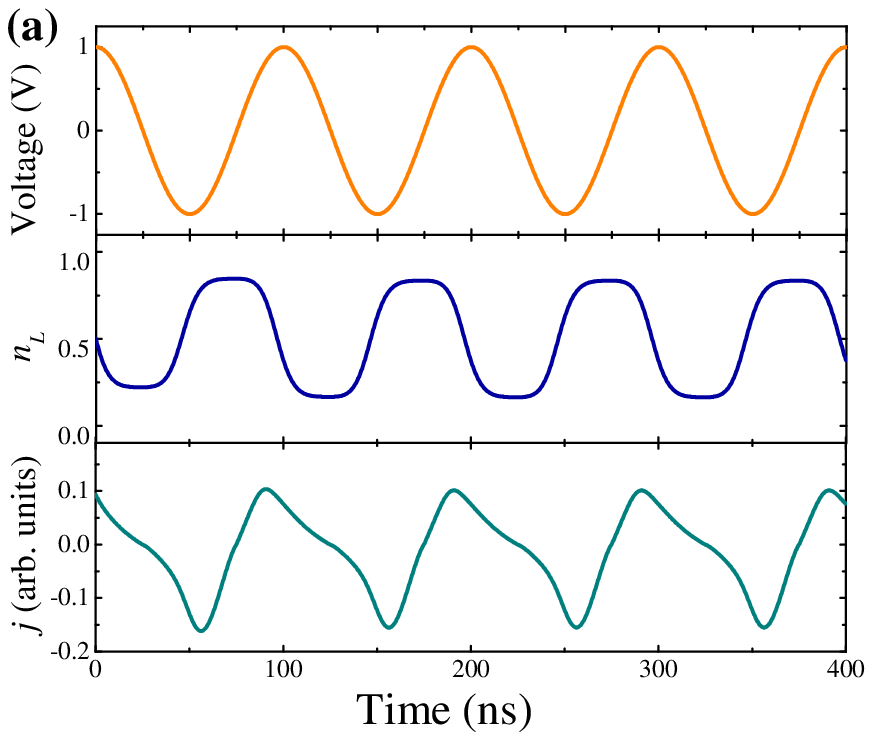}
\includegraphics[width=7cm]{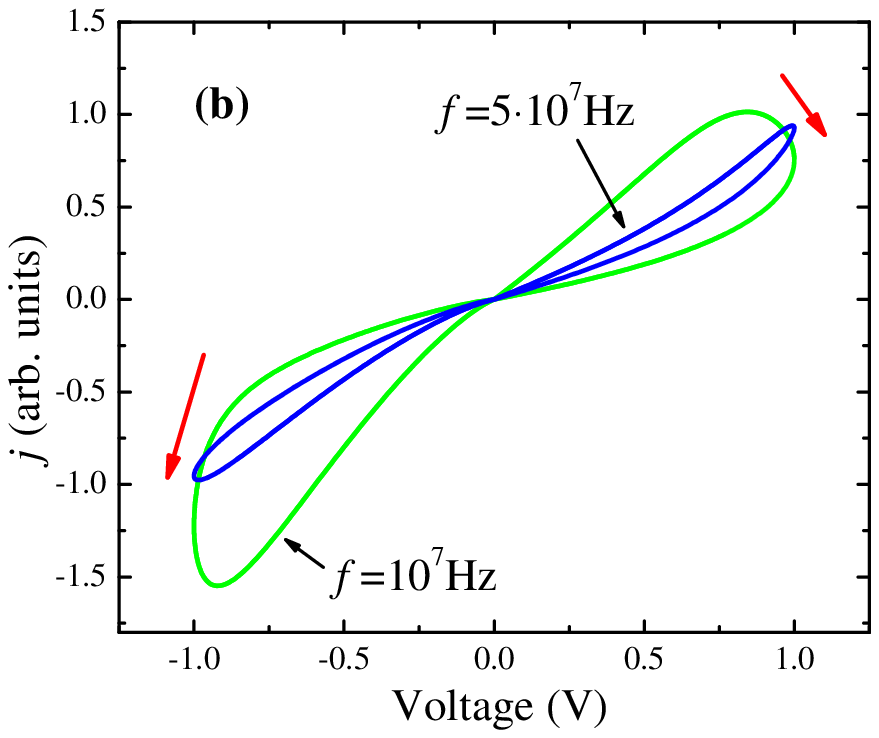}
\caption{Simulations of an AC-biased memristive device: (a) time
dependence of the applied voltage $V$, vacancies concentration
$n_L$ and current density $j$, and (b) $I-V$ curves obtained when
different frequency ac signals are applied to the device. These
plots were obtained using the following set of parameters:
$V(t)=V_0\sin(2\pi f t)$ with $V_0=1$V, $f=10^7$Hz in (a) and as
specified on the plot in (b), $E_B=1.5$eV, $T=300$K, $\xi=0.2$eV,
$E_{00}^0=2$eV, $\alpha=0.8$, $L=10$nm, $E_0=2\cdot 10^5$V/cm,
$\mu=3\cdot 10^{-10}$m$^2$/(Vs). \label{fig:devicesimulation}}
\end{figure}

We assume that the Schottky barrier width, $l$, changes linearly
with the number of the oxygen vacancies drifted to the Schottky
barrier region of the device, that is $l=l_0(1-\alpha n_L)$ where
$0<\alpha <1$ is a constant. Therefore, the maximum width of the
Schottky barrier is $l_0$ and the minimum width is
$l_0(1-\alpha)$. As the energy $E_{00}$ is inversely proportional
to $l$ (see Ref. \onlinecite{padovani66a}), we can write that
$E_{00}=E_{00}^0/(1-\alpha n_L)$ where $E_{00}^0$ is the value of
$E_{00}$ at $n_L=0$. Fig.
\ref{fig:devicesimulation}b depicts memristor's $I-V$ curves
obtained using Eq. (\ref{eq:J}). These curves have a shape
typical of the memristor pinched hysteresis loop as anticipated in
Fig.~\ref{fig:hysteresis} and typical memristor's frequency
behavior.

\section{Applications \label{Sec4}}

Potential applications of memristors are envisioned in both
digital and analog domains. Of great importance among digital
applications are non-volatile solid-state memory, signal
processing and programmable
logic~\cite{strukov07a,xia09a}. Analog applications of
memristors are based on the possibility to continuously vary their
resistance and, therefore, on their ability to store more
information than in the digital regime. Analog applications
include analog signal processing, learning circuits, programmable
analog circuits and neuromorphic circuits. Below we discuss
several analog applications of
memristors~\cite{pershin09b,pershin09c,pershin09d}.

\subsection{Learning circuits}

\begin{figure}[tb]
\centering
\includegraphics[width=8cm]{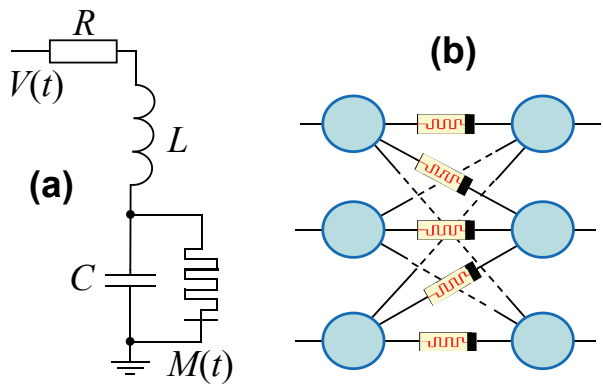}
\caption{ (a) Learning circuit~\cite{pershin09b} is composed of
four circuit elements: resistor $R$, capacitor $C$, inductor $L$
and memristor $M$. The input voltage $V(t)$ corresponds to varying
environmental conditions, the voltage on capacitor corresponds to
the response signal. For more details, see Ref.
\onlinecite{pershin09b}. (Adapted from Ref.
\onlinecite{pershin09b} with permission. Copyright American
Physical Society, 2009). (b) A hypothetical memristive neural
network involving neurons (circles) connected by memristive
synapses. \label{fig:circuits}}
\end{figure}

Quite generally, a ``learning circuit'' is an electronic circuit whose
response at a given time adapts according to signals applied to the circuit at
previous moments of time~\cite{pershin09b}. Memristors are ideal components for such
a circuit as they provide non-volatile information storage and
compatibility (as a time-dependent resistor) with other circuit
elements. Two of us (YVP and MD) have recently suggested
\cite{pershin09b} a learning circuit which mimics adaptive
behavior of the slime mold {\it Physarum polycephalum} from the
group of amoebozoa. This work was inspired by an interesting experimental
observation~\cite{saigusa08a}: when {\it Physarum polycephalum} is
exposed to a pattern of periodic environmental changes, it learns
and adapts its behavior in anticipation of the next stimulus to
come. It was demonstrated~\cite{pershin09b} that such behavior can
be described by the response of a simple electronic circuit shown
in Fig. \ref{fig:circuits}a composed of an LC contour and a memristor in parallel with the
capacitor. The memristive function employed in that work is Eq.~\ref{Mmodel2}. When a periodic signal is applied to
the learning circuit, the voltage across the capacitor significantly
changes and can exceed the threshold voltage of the memristor. This
leads to an increase in the resistance of memristor and,
consequently, in a smaller damping of the LC contour. Therefore, the LC
contour oscillations are maintained for a longer period of time in analogy with the same type of
behavior as the amoeba's when subject to periodical environment
changes.

\subsection{Neuromorphic circuits}

Neuromorphic circuits are circuits whose operation is meant to
mimic that of the brain. In these circuits, memristors can be used
as synapses whose role is to provide connections between neurons
and store information. The small size of solid-state memristors is
highly beneficial for this application since memristors' density
can be of the same order of magnitude as the density of synapses
in human brains~\cite{snider08a}. Therefore, using memristors, the
fabrication of an artificial neural network of a similar size of a
biological brain becomes possible.

Typically, an artificial neural network consists of many
artificial neurons connected by artificial synapses as we show schematically in
Fig. \ref{fig:circuits}b. The first memristor-based neural network
realized experimentally~\cite{pershin09c} contained only three
neurons connected by two synapses. However, we (YVP and MD) have shown that even such a small
network could perform a fundamental property of the human brain: associative
memory. Associative memory is closely
related to the so-called Pavlovian training~\cite{pavlov27a} in which
a particular response to a given stimulus develops. The most notable experiment
of associative memory is that of a dog to which food is shown and, at the same time,
the tone of a bell is rang so that, with time, the dog salivates at the ring of the bell only. In the
electronic 3-neuron network demonstrated in Ref.~\cite{pershin09c}, two input neurons were responsible for
the ``sight of food'' and ``sound'' events, while the output neuron
generated a ``salivation'' command. It was shown that, if one starts
with an untrained state of the synapse connecting the input ``sound''
neuron and output neuron and exposes the neural network to ``sight
of food'' and ``sound'' signals simultaneously then an association
between ``food'' and ``sound'' develops and an output signal is
generated when only the ``sound'' input signal is applied.

\subsection{Programmable analog circuits}

In programmable analog circuits, memristors can be used as digital
potentiometers~\cite{pershin09d}. The main idea is to apply small
amplitude voltages to memristors when they are used as analog
circuit elements and high amplitude voltage pulses for the purpose
of memristor's resistance programming. Since the state of
memristor appreciably changes only when the voltage applied to it
exceeds a certain threshold~\cite{strukov09a}, the resistance of
memristor is constant in the analog mode of operation and changes
by discrete values with each voltage pulse. Using this idea,
several programmable analog circuits demonstrating memristor-based
programming of threshold, gain and frequency were demonstrated
\cite{pershin09d}.

\section{Conclusions and outlook \label{Sec5}}

As this short review has shown, memory circuit elements are new promising components
for future electronics. Their advantage is based on a combination of a
history-dependent behavior with properties of a basic circuit element (such as
resistance, capacitance or inductance). Of equal importance is the possibility of
realizing memory elements at the nanoscale. The cost of certain implementations of these elements should be low
because of their simple structure. We therefore anticipate that as memristors, memcapacitors and
 meminductors possibly develop into commercially available products in the coming years, novel exciting
applications will be developed with impact not only for technology but also for fundamental science.

\section*{Acknowledgments}
This work has been
partially funded by the NSF grant No. DMR-0802830.

\newpage
\bibliography{memcapacitor}
\end{document}